\PassOptionsToPackage{dvipsnames}{xcolor}
\documentclass[sigconf,screen,authorversion]{acmart}

\usepackage[all]{nowidow}
\usepackage{xcolor}
\usepackage{subcaption}
\usepackage{hyperref}
\usepackage{acronym}

\usepackage[capitalise,noabbrev]{cleveref}
\crefformat{section}{\S#2#1#3} 
\crefformat{subsection}{\S#2#1#3}
\crefformat{subsubsection}{\S#2#1#3}

\usepackage[disable]{todonotes}
\copyrightyear{2023} 
\acmYear{2023} 
\setcopyright{acmlicensed}\acmConference[SESAME '23]{The 1st Workshop on SErverless Systems, Applications and MEthodologies }{May 8, 2023}{Rome, Italy}
\acmBooktitle{The 1st Workshop on SErverless Systems, Applications and MEthodologies (SESAME '23), May 8, 2023, Rome, Italy}
\acmPrice{15.00}
\acmDOI{10.1145/3592533.3592808}
\acmISBN{979-8-4007-0185-6/23/05}

\begin{document}

\author{Trever Schirmer}
\affiliation{%
    \institution{TU Berlin \& ECDF}
    \city{Berlin}
    \country{Germany}}
\email{ts@mcc.tu-berlin.de}

\author{Nils Japke}
\affiliation{%
    \institution{TU Berlin \& ECDF}
    \city{Berlin}
    \country{Germany}}
\email{nj@mcc.tu-berlin.de}

\author{Sofia Greten}
\affiliation{%
    \institution{TU Berlin \& ECDF}
    \city{Berlin}
    \country{Germany}}
\email{sog@mcc.tu-berlin.de}

\author{Tobias Pfandzelter}
\affiliation{%
    \institution{TU Berlin \& ECDF}
    \city{Berlin}
    \country{Germany}}
\email{tp@mcc.tu-berlin.de}

\author{David Bermbach}
\affiliation{%
    \institution{TU Berlin \& ECDF}
    \city{Berlin}
    \country{Germany}}
\email{db@mcc.tu-berlin.de}

\title[The Night Shift: Understanding Performance Variability of Cloud Serverless Platforms]{The Night Shift: Understanding Performance\\ Variability of Cloud Serverless Platforms}

\begin{abstract}
    Function-as-a-Service is a popular cloud programming model that supports developers by abstracting away most operational concerns with automatic deployment and scaling of applications.
    Due to the high level of abstraction, developers rely on the cloud platform to offer a consistent service level, as decreased performance leads to higher latency \textit{and} higher cost given the pay-per-use model.
    In this paper, we measure performance variability of Google Cloud Functions over multiple months.
    Our results show that diurnal patterns can lead to performance differences of up to 15\%, and that the frequency of unexpected cold starts increases threefold during the start of the week.
    This behavior can negatively impact researchers that conduct performance studies on cloud platforms and practitioners that run cloud applications.
\end{abstract}



\maketitle

\section{Introduction}
\label{sec:intro}

Function-as-a-Service (FaaS) is a serverless cloud computing delivery model where developers compose their applications from event-driven stateless functions and all operational tasks are managed by the cloud provider~\cite{Hendrickson2016-pw,paper_bermbach_cloud_engineering,Eismann_2021_Why}.
Functions are billed on a pay-per-use basis at second or even microsecond granularity and offer rapid elasticity and scale~\cite{Hendrickson2016-pw,paper_bermbach_cloud_engineering,Castro_2019}.
The abstraction from operational concerns has made FaaS a popular cloud execution model, with offerings by all major cloud providers, e.g., Amazon Web Services Lambda\footnote{\url{https://aws.amazon.com/lambda/}} and Google Cloud Functions\footnote{\url{https://cloud.google.com/functions/}}~\cite{paper_bermbach_cloud_engineering,Baldini_2017_Trilemma,Castro_2019}.

The flip side of high levels of resource sharing in the cloud and abstracting from resource management is that developers must rely on the cloud platform provider to offer stable and consistent performance.
Somewhat counterintuitively, FaaS users must actually pay \emph{more} when a FaaS platform underperforms and latency is higher, as billed function execution is also longer~\cite{Baldini_2017_Trilemma,Schirmer_2022_fusionizePaper}.
Most cloud FaaS platforms do not offer service level agreements beyond limited guarantees regarding general uptime~\cite{aws2022sla,gcp2021sla,zhao2022supporting}.

Cloud computing is subject to performance variations~\cite{nylander2020cloudspeculation,paper_bermbach_how_soon_is_eventual}, and FaaS is no exception, as previous studies on the long-term (day-to-day) performance changes of FaaS platforms have shown~\cite{Eismann_2022_stability,vaneyk2020specrg,Lambion_2022_AWSDaily}.
A general improvement of FaaS services over time is expected as platform providers update and advance there infrastructure~\cite{Werner_2022_Hardless}, yet there are also much finer effects in the short term.

In this paper, we benchmark and analyze these effects using highly frequent (every 40s) cloud FaaS benchmarks over the course of two months against Google Cloud Functions (GCF), which has received fewer attention in previous studies.
Our main finding is that performance varies greatly during the course of the day, with an increase of request-response latency of up to 15\% and more than three times as many unexpected cold starts from day to night \emph{within the same day}, and that these effects are most noticeable at the start of the week.
These findings impact how we interpret the results of cloud FaaS benchmarks and are significant for cloud FaaS developers.
In summary, we make the following contributions:

\begin{itemize}
    \item Based on a number of existing serverless benchmarks, we propose a methodology for evaluating temporal performance variations in cloud FaaS platforms (\cref{sec:methodology}).
    \item We execute our benchmark with frequent (every 40s) runs on Google Cloud Functions over the course of two months and show how request-response latency exhibits strong (up to 15\% difference) diurnal variation (\cref{sec:eval:performance}).
    \item We further uncover an increase in unexpected cold starts at specific times of the day that suggest increased rates of instance recycling in GCF during times of high demand (\cref{sec:eval:coldstarts}).
    \item We evaluate long-term trends in our data and identify possible causes (\cref{sec:eval:longterm}).
    \item We survey existing cloud FaaS benchmarking studies and find that almost two thirds provide insufficient information on how the performance variability effects we identify are controlled for (\cref{sec:impli:benchmarks}).
    \item We discuss implications of our findings for practitioners that run applications on cloud FaaS platforms (\cref{sec:impli:applications}).
\end{itemize}

In order to enable other researchers and practitioners to extend and replicate our experiments, we make the artifacts used to produce this paper available as open-source\footnote{\url{https://github.com/umbrellerde/night-shift-code}}.
\section{Related Work}
\label{sec:related}
Existing research on FaaS performance variability focuses on \emph{either} short-term \emph{or} long-term (i.e., more than a week) variability.
Short-term studies include a report by Lambion et al.~\cite{Lambion_2022_AWSDaily}, who have benchmarked the performance of various functions on AWS Lambda over the course of a day.
The authors execute their functions in different time zones and using different hardware architectures, and find a 6\% shorter function duration during the night. 
Mahmoudi et al.~\cite{mahmoudi2021temporal} propose an analytical performance model to predict performance metrics of functions. 
To validate their model, they repeat the same one-hour experiment ten times, and find that request arrival rate, average response times, and function timeout can be used to predict performance up to five minutes in advance.
Ginzburg and Freedman~\cite{ginzburg2020serverless} analyze performance variations on AWS Lambda over the course a week. 
They call 1000 functions every two hours and show that the daily performance of the same function inside the same region and between regions can vary significantly, which is mainly caused by local inactivity and lack of performance isolation between tenants.
Since they only measure for one week, their analysis focuses on daily variations, which they measure at 1-2\%.

A long-term study of serverless systems is presented by Eismann et al.~\cite{Eismann_2022_stability}, who have executed the same serverless application on Lambda once a day over ten months. 
They find long-term performance changes that are likely to be caused by platform changes, and short-term variations between days.
Figelia et al.~\cite{figiela2018performance} measure the performance of various functions running on Lambda over seven months, but do not analyze their dataset for regular variability.

The focus of our paper is to close the gap between long-term and short-term studies by collecting frequent measurements over a longer period of time.
Additionally, we go beyond the focus on AWS Lambda and present a general methodology for experiments that are applicable to all FaaS platforms.
We collect our results on Google Cloud Functions, which has received fewer attention in previous studies despite its popularity.
\section{Methodology}
\label{sec:methodology}

To assess performance variations in FaaS, we repeatedly execute a FaaS function in short intervals across a large time span.
By controlling for execution region, resource parameters (through the memory option on cloud FaaS providers), and function type, we can focus on cloud platform performance.

\subsubsection*{Functions}

We use three FaaS functions from existing serverless benchmarks in our experiments.
All functions perform isolated computations that do not rely on external services, the performance of which could influence our results~\cite{paper_grambow_befaas}.
The \emph{float} workload of Kim et al.~\cite{Kim_2019_FunctionBench} that performs floating point operations.
The \emph{matrix} function of Werner et al.~\cite{Werner_2018_Matrix} that performs matrix multiplication.
Finally, we adapt the face detection model of Barosum et al.~\cite{Barsoum_2016_onnx} for the \emph{ml} function.
To minimize the impact of external fluctuations on our measurements, we embed all inputs directly into the functions.

The resources available to a function instance are determined by the memory configured for that function:
On GCF, the amount of vCPUs allocated to a function instance is tiered and increases with every multiple of 128MB memory, while AWS Lambda scales vCPUs linearly with memory~\cite{Cordingly_2022_memorysizes}.
To capture effects of resource configurations, we deploy \textit{float} and \textit{matrix} with 128MB, 256MB, and 512MB of memory, while \textit{ml} is deployed with 512MB and 1024MB as it has higher resource requirements.

\subsubsection*{Execution}

Cloud function invocations can be both ``warm'' and ``cold'':
When a function is invoked for the first time, a new function instance is created.
This is called a ``cold start'' and incurs a creation overhead~\cite{Manner_2018_Coldstarts,Bardsley_2018_coldstarts,paper_bermbach_faas_coldstarts}.
Subsequent (but possibly not parallel) invocations of the same function can reuse the existing instance and avoid this overhead, the ``warm starts''.
Typically, cloud platforms will keep existing instances for future invocations for a limited amount of time and then evict them to reclaim resources~\cite{copik2021sebs}.

To capture both cold and warm start latencies, we invoke functions in loops:
We first call the function once, creating a cold start, and then call the function again.
In theory, this second invocation should be served by the existing function instance.
We then wait 20 minutes to make sure that the next function call is a cold start again and restart the same loop.
To collect more measurement points than twice every 20 minutes, we deploy parallel copies of a function that we cycle through.

\subsubsection*{Metrics}

We consider three main metrics: request-response latency, unexpected cold starts, and long-term trends.
For request-response latency, we use the billed duration that is output for every function execution by the FaaS platform.
Unexpected cold starts are cold starts that occur directly after a function has already been called once, so that they should be warm.
Unexpected cold starts imply that a platform was unable to find a warm function instance, possibly because it has been evicted due to resource contention.

Finally, we conduct a seasonal trend decomposition using LOESS (STL)~\cite{cleveland1990stl}, which can handle complex seasonal patterns.
We use the following model for our data:

\begin{equation*}
    y_t = T_t + S_t + I_t
\end{equation*}

, where $T_t$ represents the trend, $S_t$ represents a seasonal component (days in our case), and $I_t$ represents the remaining noise.
We fit our data to this model using a method to create smoothed estimates with a seasonality of one day.
The trend component $T_t$ shows the overall progress in the billed duration over the whole duration of the experiment.
A non-flat trend line indicates that longer-term changes to the platform have occurred, e.g., long-term seasonal changes or updates to the platform that influenced performance.
The seasonal component $S_t$ shows periodic recurring deviations in the data from the trend.
The $I_t$ component is random noise centered around 0.
Outliers in the noise indicate that the performance at specific times could not be explained by the previous two components.
The trend component from the STL is also used for Change Point Detection~\cite{Truong_2020_ChangePoint}, which can detect structural changes in data.

\section{Experiments}
\label{sec:eval}

In this section, we present initial results of our experiments with Google Cloud Functions in the \texttt{europe-west3} region.
The measurement period started on Dec 12, 2022 and ended on Feb 27, 2023.
We report all execution times in local time (CET).
We first analyze the performance variability of the platform (\cref{sec:eval:performance}).
Afterwards, we explore unexpected cold starts (\cref{sec:eval:coldstarts}) and outliers as well as change points and long-term trends in our data (\cref{sec:eval:longterm}).

\subsection{Performance Variability}
\label{sec:eval:performance}

\begin{figure}
    \centering
    \includegraphics[width=\linewidth]{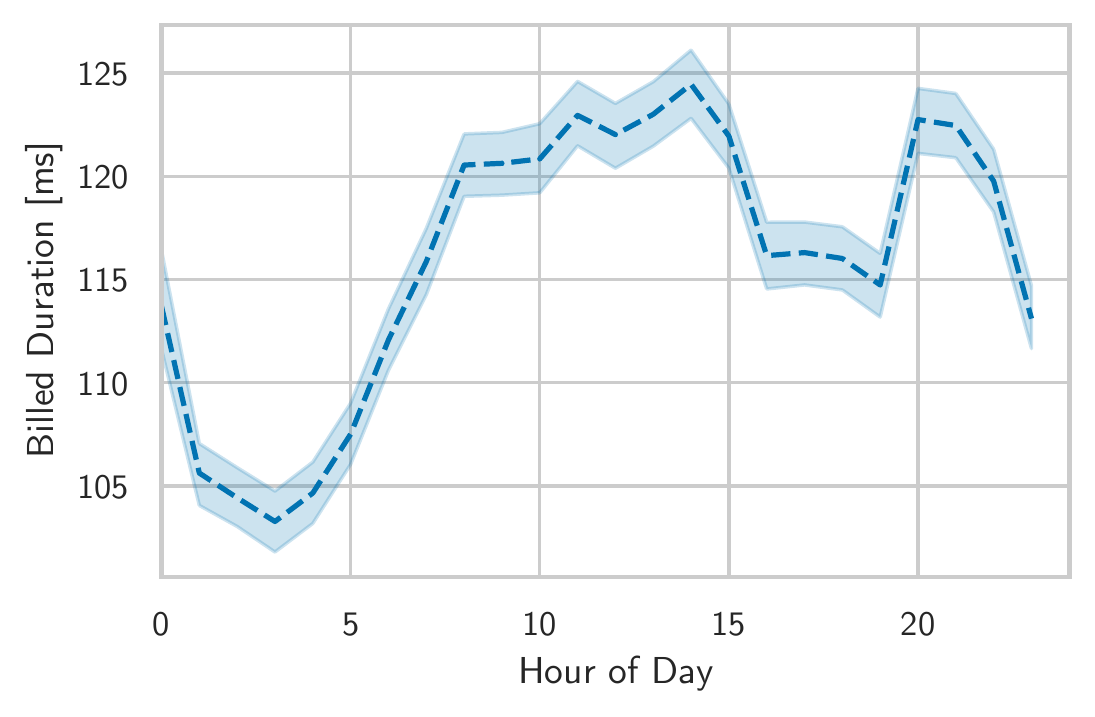}
    \caption{Billed Duration of warm calls to \textit{float} with 128MB memory.
        During working hours, the billed duration increases by up to 15\%.
        The area around the dashed line shows the 95\% confidence interval.}
    \Description{A line-plot that clearly dips during the hours 0-7 (<115ms), then has a peak around 7-15 (>118ms). Between 15-20, the performance ist constant around 115ms, and the last peak ist at 20 at around 120ms. The error bars are always between 2-4ms.}
    \label{fig:eval:floatHod}
\end{figure}

To analyze performance variability, we analyze the billed duration of comparable invocations.
We show the billed duration of the \textit{float} function with 128MB memory in \cref{fig:eval:floatHod}.
We observe a clear performance increase during the night, with a noticeable latency spike during working hours.
The average billed duration between 23:00 and 06:00 was 106ms, and increased by 15\% to 122ms between 07:00 and 16:00.
When aggregated by the day of the week, the average billed duration fluctuates between 113.86ms on Saturdays and 117.28ms on Mondays.
Overall, billed duration is slightly lower on the average weekend compared to the start of the week.
The weekly trend is much smaller than the daily trend, as the billed duration only decreases by \textasciitilde4\% during the weekend.

\begin{figure}
    \centering
    \begin{subfigure}{0.5\textwidth}
        \includegraphics[width=\linewidth]{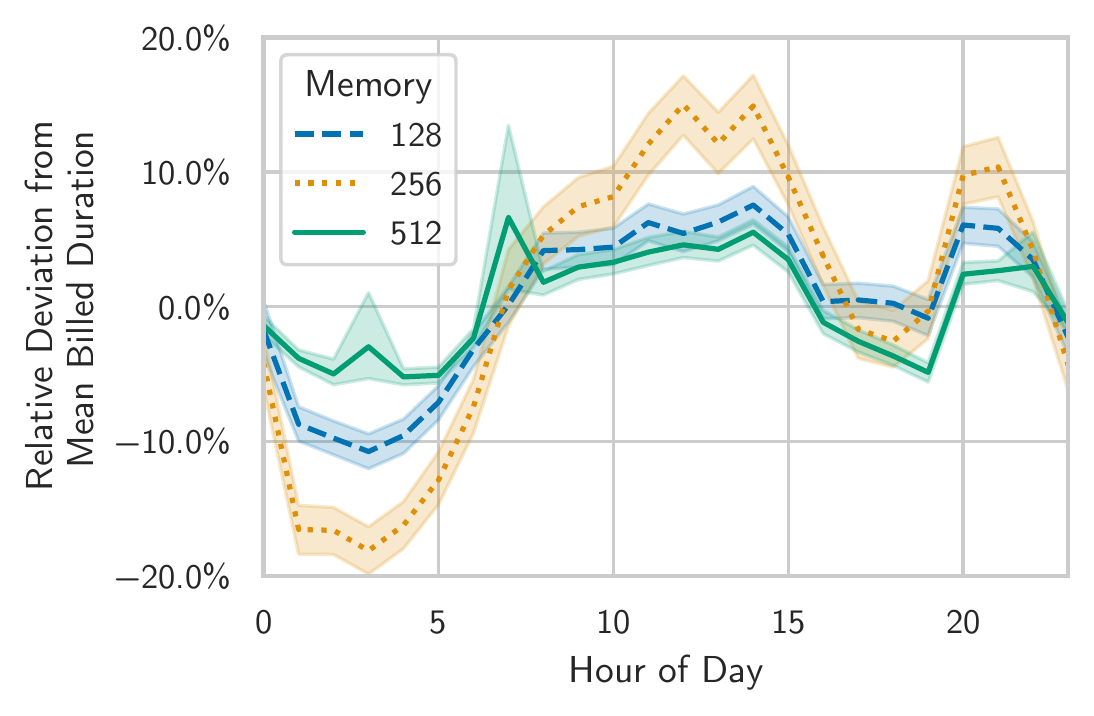}
        \subcaption{\textit{float}}
        \Description{A line chart that shows the same pattern as \cref{fig:eval:floatHod}. The amplitude is most pronounced with the 256MB-line, which dips to -20\% between 0-5 and increases to 15\% between 10-15. The 128MB line is less pronounced (-10\% and +5\% respectively), and the 512MB line is always within 5\% of its average.}
    \end{subfigure}
    \begin{subfigure}{0.5\textwidth}
        \includegraphics[width=\linewidth]{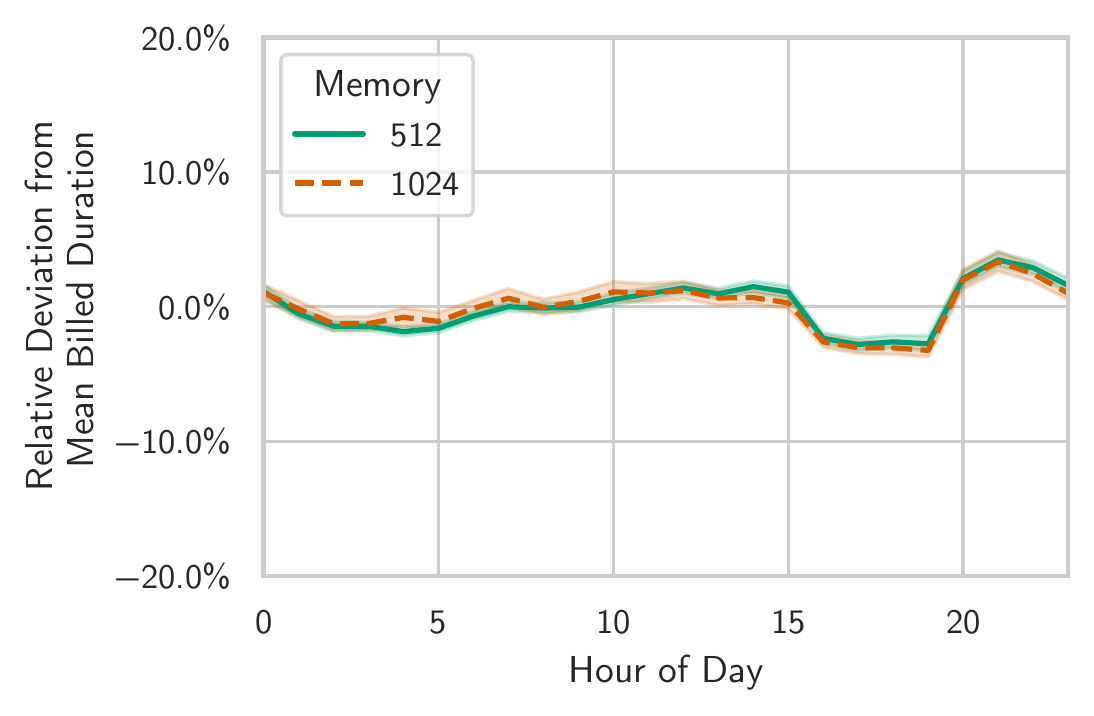}
        \subcaption{\textit{ml}}
    \end{subfigure}
    \caption{Performance change over a day of warm instances ordered by memory size. The y-Axis is normalized to the average and shows the relative change, e.g., bigger values show a bigger deviation from the average billed duration. The area around the curves shows the 95\% confidence interval.}
    \Description{The 512MB and 1024MB lines follow the daily pattern, but the variance within a day is <2\% from 0-15 and <4\% afterwards.}
    \label{fig:eval:relChange}
\end{figure}

\begin{figure}
    \centering
    \includegraphics[width=\linewidth]{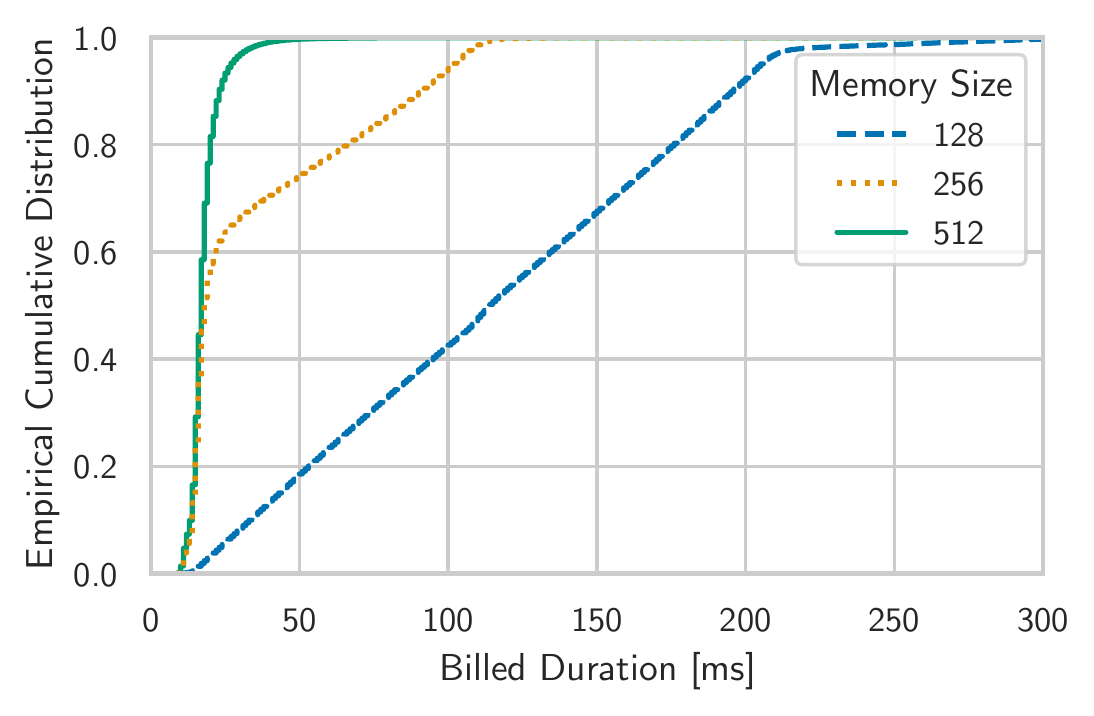}
    \caption{Cumulative distribution of billed durations of the float function without cold starts. We argue that GCF uses 128MB and 512MB containers to execute 256MB functions.}
    \Description{The billed duration is between \textasciitilde{}20ms and 250ms for most calls. The 128MB calls form a straight line with 90\% taking less than \textasciitilde 200ms, and 512MB follows a normal distribution around 25ms. The 256MB-line first follows the 512MB line up until 50\%. Afterwards it only increases with the same angle as the 128MB line (i.e., they run in parallel).}
    \label{fig:eval:floatSizeEcdf}
\end{figure}

When looking at larger memory sizes, the relative performance change over time becomes smaller.
As shown in \cref{fig:eval:relChange}, the \textit{float} function with 128MB memory changes \textasciitilde10\% during a day, while 512MB only changes up to 5\%.
The \textit{ml} functions with 512MB and 1024MB memory equally changed \textasciitilde4\% during an average day.
Noticeably, the average billed duration of the \textit{float} function with 256MB of memory differed more than 15\% during a day.
This can be explained by looking at the distribution of latency values over all invocations (\Cref{fig:eval:floatSizeEcdf}):
Around 50\% of functions with a memory size of 256MB follow the same distribution as functions with 512MB of memory, and the other half follows the same distribution as the 128MB functions.
The results for the \textit{matrix} function exhibit similar results but are omitted due to space constraints.
The high variability in performance and uneven distribution of billed durations indicates that GCF internally uses 128MB and 512MB function instances to handle requests to the 256MB functions, as also shown by Malawski et al.~\cite{malawski_2018_gcfopportun}.

While cold start durations also follow the daily patterns shown in \cref{sec:eval:performance}, the configured memory size has no impact on the duration of cold starts.
The billed durations of cold starts follow a normal distribution and are on average around 9-10x longer than the average warm latency, but all memory sizes follow the same distribution.
This indicates that the cold start overhead is dependent on resources that can be configured by changing the function configuration.

\subsection{Unexpected Cold Starts}
\label{sec:eval:coldstarts}

\begin{figure}
    \centering
    \includegraphics[width=\linewidth]{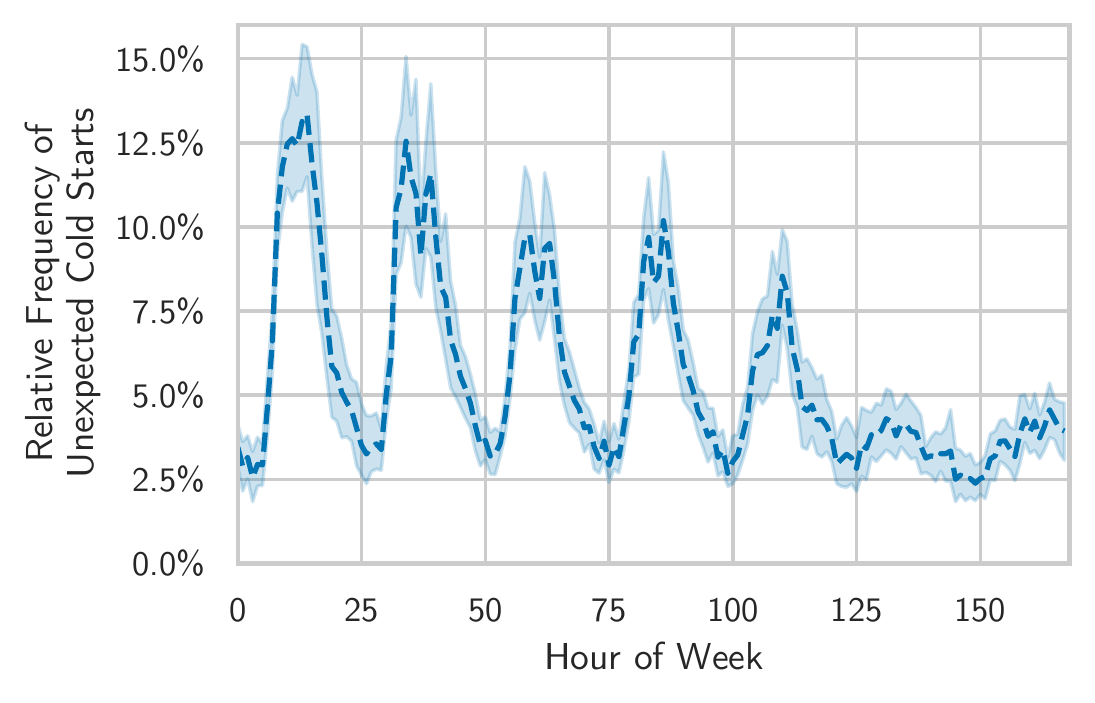}
    \caption{Relative frequency of unexpected cold starts, averaged by the hour of the week they happened in. On Mondays during the day, up to 13\% of invocations can be unexpected cold starts, compared to less than 5\% during the night and on weekends. The area around the curve shows the 95\% confidence interval.}
    \Description{Line plot showing regular patterns over the week and every day. During every day, there are more cold starts during the day than night hours. While during the night, cold starts are always around 2.5-3\%, they increase up to 12\% on Monday. The peaks decrease during the week with a peak of only 7.5\% on Friday. On the weekend, unexpected cold starts are always lower than 5\%.}
    \label{fig:eval:unexpectedCold}
\end{figure}

We show the relative frequency of unexpected cold starts in \cref{fig:eval:unexpectedCold}.
While the billed duration of warm instances seems to only follow a daily trend, the frequency of unexpected cold starts has a weekly seasonality, with clear trends of increased cold starts during working hours.
On average, the frequency of unexpected cold starts was 3.7\% during the night (20:00—08:00), 3.6\% during the weekend, 9.8\% during working hours (09:00—17:00 Mon—Fri), and 12.3\% during working hours on Monday.
For comparison, there were less than 0.15\% unexpected warm starts, where a function instance was still warm after more than 20 minutes.

\subsection{Outliers \& Long-Term Trend}
\label{sec:eval:longterm}

\begin{figure}
    \centering
    \includegraphics[width=\linewidth]{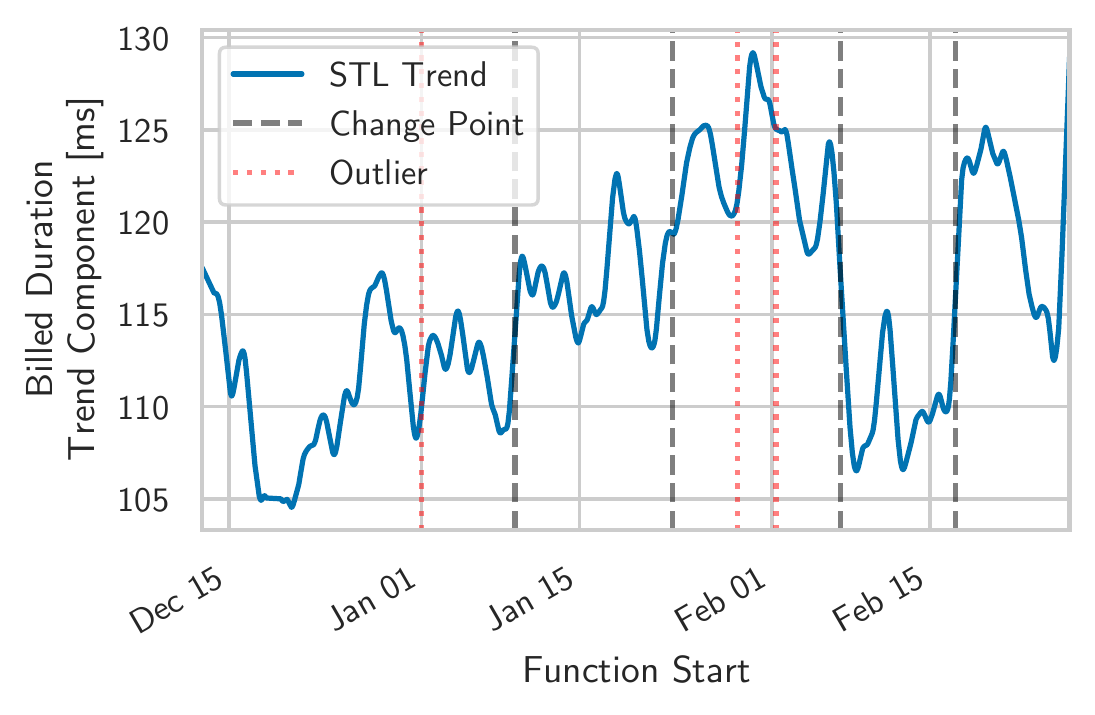}
    \caption{Trend Component of the STL, with Change Points and Outliers marked as horizontal lines.}
    \Description{Line Plot showing a trend that is between 105ms and 130ms, between Dec 15, 2022 and Feb 15, 2023. The trend seems to steadily increase from Dec 20 (105ms) to Feb 01 (125ms), after which it decreases until Feb 8 (110ms), holds steady until Feb 15, after which it increases back to \textasciitilde{}120ms. Change points are on 2023-01-09T07:00:00, 2023-01-23T06:00:00, 2023-02-07T01:00:00, 2023-02-17T06:00:00. Outliers are on 2023-01-01 00:00:00,2023-01-03 06:00:00, 2023-01-27 03:00:00, 2023-01-29 00:00:00, 2023-02-01 09:00:00}
    \label{fig:lt:changePoint}
\end{figure}

Based on the STL introduced in \cref{sec:methodology}, we show change points and outliers in \cref{fig:lt:changePoint}.
We define an outlier as every hour during which the average execution duration is outside the fourth interquartile range, i.e., more than four times the difference between the first and third quartile, away from the average.
Our data contains four outliers, which were all within three days of the turn of the month.
This indicates to us that the platform is under unusual load at these times, possibly due to additional load from monthly jobs.

All change points, i.e., points when the average execution duration changed, occurred during the night, indicating that they coincide with scheduled updates to the platform.

Over our whole measurement period, there is no clear permanent trend towards better or worse overall performance.
Based on the long-term study by Eismann et al.~\cite{Eismann_2022_stability}, we only expect to find permanent trends in longer measurement periods.
Compared to the authors' study of AWS, which finds statistical trends below 10\%, our trend component changes between 105ms during December to 128ms during February, a 21\% increase.

\section{Implications}
\label{sec:impli}

In this section, we discuss how our findings impact serverless systems.
First, we focus on the implications for benchmarking.
Afterwards, we describe implications on serverless applications.

\subsection{Validity of Benchmarks}
\label{sec:impli:benchmarks}

\begin{figure}
    \centering
    \includegraphics[width=\linewidth]{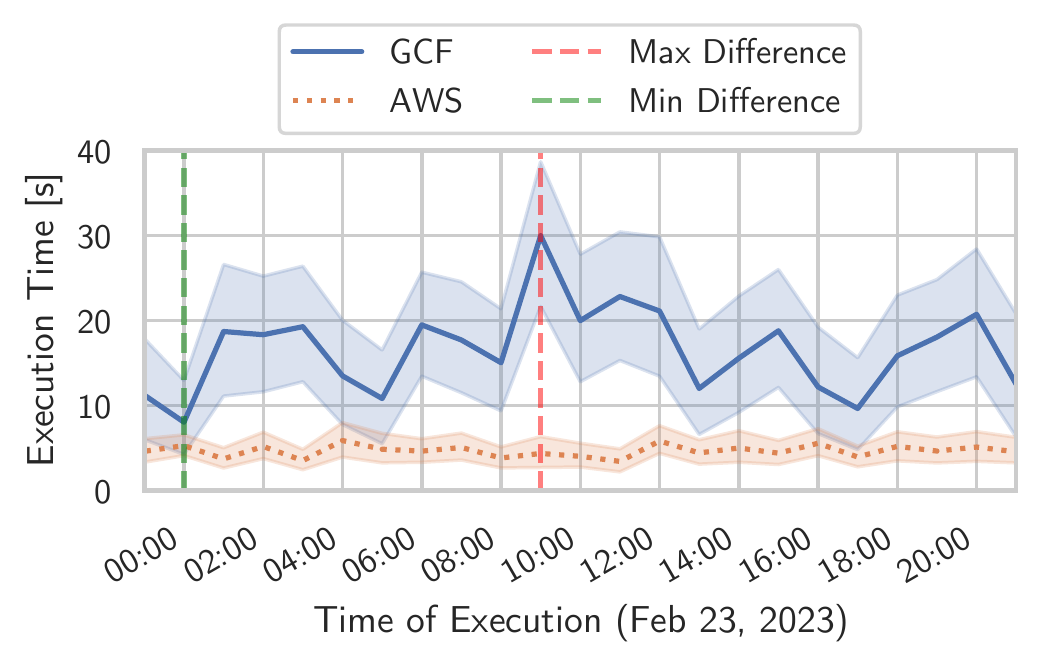}
    \caption{Execution Times of the \texttt{110\-.dynamic\--html} Benchmark on GCF and Lambda (excerpt). The minimum and maximum measured difference of the whole experiment are marked as horizontal lines.}
    \Description{Line plot showing Feb 23, 2023 between 00:00 and 20:00. The AWS Line is constant at around 5s, while GCF varies between 10s and 30s. At the time of minimal difference, AWS is ever so slightly slower than usual while GCF is unusually fast, and during the maximum difference, GCF is unusually slow. GCF follows the faily patterns from \cref{fig:eval:floatHod}, which high variance.}
    \label{fig:eval:sebsConsistency}
\end{figure}

When running benchmarks, researchers want to minimize effects of external factors to their measurements, which otherwise might confound results~\cite{book_bermbach2017_cloud_service_benchmarking}.
In the case of benchmarking serverless systems, we have shown that the number of cold starts and the performance of functions undergoes changes within a single day.
For comparative performance studies, benchmark results are only comparable if experiments are conducted at comparable times.
If an experiment is short, i.e., does not capture the performance variation of an entire day, daily variations can skew results.
Similarly, long-term performance changes can impact measurements taken over longer periods of time, e.g., a three-day study with one group benchmarked over the weekend and another benchmarked during the week.
A possible remediation for such experiments is adopting parallel benchmarking techniques such as duet benchmarking~\cite{bulej2020duet}.

We show an overview of existing publications on cloud FaaS performance measurements surveyed for this paper in \cref{tab:bench}.
For every paper, we give an overview of cloud platforms under test, cloud regions, and time of day of the benchmark execution, if stated.
Based on this metadata, we must assume that published results could be affected by daily performance variations if execution time is not given or different regions are used (implying different time zones).
Overall, 10 out of a total of 16 papers do not provide sufficient information to rule out effects of performance variability.
While this does not mean that the reported results are invalid, it shows that the research community has not paid enough attention to these effects in performance measurements.

\begin{table*}
    \centering
    \caption{Selection of serverless benchmarks and whether they \emph{could} be impacted by short-term performance fluctuations.
        \textbf{Bold} lines are papers where a daily impact could not be ruled out.
    }
    \label{tab:bench}

    \resizebox{0.9\textwidth}{!}{
        \begin{tabular}{@{}rccc@{}}
            \toprule

            Authors                                                    & Cloud Platform(s)             & Region(s) Specified                & Execution Time of Day       \\

            \midrule

            \textbf{Copik et al.~\cite{copik2021sebs}}                 & \textbf{AWS, Azure, GCP}      & \textbf{US (AWS), EU (Azure, GCP)} & \textbf{--}                 \\
            Eismann et al.~\cite{Eismann_2022_stability}               & AWS                           & --                                 & daily (19:00) for 10 months \\
            \textbf{Figiela et al.~\cite{figiela2018performance}}      & \textbf{AWS, Azure, IBM, GCP} & \textbf{EU (AWS), US (GCP)}        & \textbf{every 5 minutes}    \\
            Jackson et al.~\cite{jackson2018investigation}             & AWS, Azure                    & --                                 & hourly over 6 days          \\
            \textbf{Grambow et al.~\cite{paper_grambow_befaas}}        & \textbf{AWS, Azure, GCP}      & \textbf{EU}                        & \textbf{--}                 \\
            \textbf{Kim et al.~\cite{Kim_2019_FunctionBench}}          & \textbf{AWS, Azure, GCP}      & \textbf{--}                        & \textbf{--}                 \\
            \textbf{L{\'o}pez et al.~\cite{lopez2018comparison}}       & \textbf{AWS, Azure, IBM}      & \textbf{--}                        & \textbf{--}                 \\
            Malawski et al.~\cite{malawski_2018_gcfopportun}           & AWS, GCP                      & EU (AWS), US (GCP)                 & ``permanently''             \\
            Manner et al.~\cite{Manner_2018_Coldstarts}                & AWS, Azure                    & --                                 & --                          \\
            \textbf{McGrath et al.~\cite{mcgrath2017serverless}}       & \textbf{AWS, Azure, GCP}      & \textbf{--}                        & \textbf{--}                 \\
            Pelle et al.~\cite{pelle2019towards}                       & AWS                           & ``multiple regions''               & \textbf{--}                 \\
            \textbf{Scheuner et al.~\cite{Scheuner_2022_TriggerBench}} & \textbf{AWS}                  & \textbf{US}                        & \textbf{--}                 \\
            Shahrad et al.~\cite{Shahrad_2020}                         & Azure                         & ``entire infrastructure''          & every minute                \\
            \textbf{Somu et al.~\cite{Somu_2020_PanOpticon}}           & \textbf{AWS, GCP}             & \textbf{--}                        & \textbf{--}                 \\
            \textbf{Werner et al.~\cite{Werner_2018_Matrix}}           & \textbf{AWS}                  & \textbf{EU}                        & \textbf{--}                 \\
            \textbf{Zhang et al.~\cite{zhang2019video}}                & \textbf{AWS, GCP}             & \textbf{--}                        & \textbf{--}                 \\

            \bottomrule
        \end{tabular}
    }

\end{table*}

As an example, we replicate an experiment from Copik et al.~\cite{copik2021sebs} that compares the execution time of a dynamic HTML generator (\texttt{110.dynamic\--html}) between AWS Lambda and GCF.
We deploy this function in the \texttt{eu\--central\--1} (Lambda) and \texttt{europe\--west3} (GCF) regions with 128MB of memory and run 50 sequential invocations every hour over the course of five days.
As shown in \cref{fig:eval:sebsConsistency}, the performance on GCF exhibits temporal variations that can skew results:
The smallest performance difference during our experiments happened on the 23rd of February 2023 at midnight, when the average performance difference was <2.8s (GCF 52\% slower).
Shortly after, at 09:00, we observe the largest performance difference with the average difference increasing to >25s (GCF 6.8$\times$~slower).
These performance changes show that short-term variation in performance between serverless platforms can have a significant impact on benchmarking results and need to be controlled for.
We recommend repeating experiments over the course of a day and mentioning execution time when describing experiment setup.

\subsection{Application Performance}
\label{sec:impli:applications}

The performance variability that we have shown for GCF affects serverless applications in several ways:
During daytime, functions have increased latency, suffer more cold starts, and their execution cost is increased due to the pay-by-second billing model.
For low-latency, event-driven functions, it is not feasible to postpone their execution to the night or a weekend to decrease costs.
A possible way forward, however, is to shift function execution to another cloud region with better performance.
This may increase network latency and transmission costs, but an up to 20\% reduction on function execution times and the associated decrease in costs can outweigh this overhead for long(er)-running functions.
Such an approach requires constant evaluation of FaaS platform performance in different regions, possibly also based on application metrics~\cite{paper_bermbach_s3_longterm_study}.

Researchers have also proposed systems that adapt FaaS applications to improve performance and cost on cloud FaaS platforms.
Such systems rely on initial performance measurements of applications on cloud platforms~\cite{Elgamal_2018,Czentye_2019,Horovitz_2019_VmMlFaaS,Cordingly_2022_memorysizes} or a feedback loop between platform, application, and optimizer~\cite{Schirmer_2022_fusionizePaper}.
Both approaches are affected by performance variability of the FaaS platform, as the optimizer or model cannot differentiate between performance changes that are caused by deployment updates and those that are caused by platform instability.
A possible way forward for these systems is to control for temporal performance variations, e.g., by deploying multiple parameter sets concurrently or performing longer initial measurements.
\section{Limitations \& Future Work}
\label{sec:limits}

We have shown considerable performance variability in Google Cloud Functions and discussed how these affect applications and performance measurement research.
We plan to build on this initial work in the future to arrive at a more holistic view of performance variability in cloud FaaS platforms.

\noindent
\textit{FaaS Platforms. }
Our initial experiments are limited mostly to GCF, with some additional validation on AWS Lambda.
Although we have seen that in our experiments, Lambda suffers from less performance variability than GCF, parameters such as memory size, hardware architecture, geographical region, or programming language could further influence variability.
We plan to conduct additional experiments on different FaaS platforms in the future, controlling for these additional parameters.

\noindent
\textit{Platform Changes. }
FaaS platforms are evolving quickly, and our measurements and experiments can only capture the behavior of such a platform at a specific point in time.
Continuous updates could increase or even eliminate the performance variability effects we observe in the future, making continuous measurements important.
Subsequently, researchers that want to account for the described behavior in their own measurements on FaaS platforms should conduct their own experiments using our methodology, as our measurement results may be outdated by then.

\noindent
\textit{Performance Dimensions. }
The functions we use in our experiments are CPU-bound, which gives a good indication for general platform performance and minimizes the impact of the performance of external services.
Beyond CPU performance, other resource metrics such as memory access, disk I/O, and network latency or bandwidth can be affected by platform variability.
As our findings may not be unconditionally applicable to workloads that are bound in these dimensions, we will investigate their variation with additional functions in the future.

\section{Conclusion}
\label{sec:concl}

In this paper, we have presented the results of our multi-month performance variability benchmark measuring the performance of multiple functions on Google Cloud Functions every 40s.
Our results show that the execution duration of a function varies up to 15\% per day, and the frequency of unexpected cold starts varies over a week and per day.
While more resources reduce daily performance variability, they do not shorten cold start durations.
By looking at the long-term trend, we identify likely updates to the platform and outlier behavior around the turn of the month.
These results have implications for both researchers and practitioners.

\balance
\bibliographystyle{ACM-Reference-Format}
\bibliography{bibliography.bib}

\end{document}